\documentstyle[12pt]{article}
\begin{document}
\def\bbox{{\,\lower0.9pt\vbox{\hrule \hbox{\vrule height 0.2 cm
\hskip 0.2 cm \vrule height 0.2 cm}\hrule}\,}}
\def\laq{\ \raise 0.4ex\hbox{$<$}\kern -0.8em\lower 0.62
ex\hbox{$\sim$}\ }
\def\gaq{\ \raise 0.4ex\hbox{$>$}\kern -0.7em\lower 0.62
ex\hbox{$\sim$}\ }


\def\a{\alpha}
\def\b{\beta}
\def\g{\gamma}
\def\G{\Gamma}
\def\d{\delta}
\def\D{\Delta}
\def\e{\epsilon}
\def\h{\hbar}
\def\ve{\varepsilon}
\def\z{\zeta}
\def\t{\theta}
\def\vt{\vartheta}
\def\r{\rho}
\def\vr{\varrho}
\def\k{\kappa}
\def\l{\lambda}
\def\L{\Lambda}
\def\m{\mu}
\def\n{\nu}
\def\o{\omega}
\def\O{\Omega}
\def\s{\sigma}
\def\vs{\varsigma}
\def\S{\Sigma}
\def\vphi{\varphi}
\def\av#1{\langle#1\rangle}
\def\pa{\partial}
\def\na{\nabla}
\def\hg{\hat g}
\def\un{\underline}
\def\ov{\overline}
\def\cF{{{\cal F}}}
\def\cG{{{\cal G}}}
\def\Hsl{H \hskip-8pt /}
\def\Fsl{F \hskip-6pt /}
\def\cFsl{\cF \hskip-5pt /}
\def\ksl{k \hskip-6pt /}
\def\pasl{\pa \hskip-6pt /}
\def\tr{{\rm tr}}
\def\tcF{{\tilde{{\cal F}_2}}}
\def\tg{{\tilde g}}
\def\tphi{{\tilde\phi}}
\def\shalf{\frac{1}{2}}
\def\nn{\nonumber \\}
\def\w{\wedge}
\def\ra{\rightarrow}
\def\la{\leftarrow}
\def\be{\begin{equation}}
\def\ee{\end{equation}}
\newcommand{\brr}{\begin{eqnarray}}
\newcommand{\err}{\end{eqnarray}}

\begin{titlepage}
\setcounter{page}{0}

\begin{flushright}
COLO-HEP-424 \\
hep-th/0002174 \\
February 2000
\end{flushright}

\vspace{5 mm}
\begin{center}
{\large Brane World Scenarios and the Cosmological Constant}
\vspace{10 mm}

{\large S. P. de Alwis\footnote{e-mail: 
dealwis@pizero.colorado.edu}}\\
{\em Department of Physics, Box 390,
University of Colorado, Boulder, CO 80309.}\\
\vspace{5 mm}
\end{center}

\vspace{10 mm}

\centerline{{\bf{Abstract}}}
Brane world scenarios offer a  way of ensuring that a Poincare invariant  four dimensional world can emerge, without fine tuning, as a solution to the equations of motion of an effective action.  We discuss the different ways in which this happens, and point out  that the underlying reason is that there is a contribution to the  effective  cosmological constant which is  a constant of integration, that maybe adjusted to ensure a flat space solution. Basically this is an old idea revived in a new context and we speculate that there may be string scenarios that provide a concrete realization of it. Finally we discuss to what extent this is a solution to the cosmological constant problem.

\end{titlepage}
\newpage
\renewcommand{\thefootnote}{\arabic{footnote}}
\setcounter{footnote}{0}
\section{Introduction}

Brane world scenarios are based on the hypothesis that the three space dimensional world that we appear to be living in is a brane that is embedded
in a higher dimensional world\footnote{This is an old idea \cite{Rub} that has been revived recently in a string inspired context in \cite{jl}\cite{AADD}.}. Most of the work on this has been of a phenomenological  nature and not many attempts have been made to justify the postulates within a well defined framework for (higher dimensional) quantum gravity such as string theory. Nevertheless this activity is ``string inspired", in that an obvious candidate for such
a world is a collection   of  D-branes on which (at least in principle) the standard model can live. For most of this paper
we will not worry about a string realization  though towards the end we will suggest some possibilities.

The main issue that we are concerned with here, is that of
obtaining flat  3+1 dimensional solutions to the equations of
the effective higher dimensional theory in a natural way (i.e. without fine tuning). We will show that there are situations where   flat brane solutions can be obtained by choice of integration constant\footnote{After this work was substantially completed several papers appeared which obtain a one flat brane solution by choice of integration constant \cite{BG}\cite{ADS}\cite{KSS}. We will comment on these works in the conclusion.}. In this respect this mechanism is a realization whithin the brane world 
context of an old idea going back to \cite{OS}\cite{DvN},\cite{TAN},\cite{Witten},\cite{Hawking},\cite{Rub2}\footnote{There are also unimodular gravity scenarios that appear to have been first discussed in \cite{Ng}and are also discussed in \cite{weinberg}. However these do not fit in naturally in a string picture.}. To set the stage for the 
brane world calculations 
we will first review this argument.

Consider an effective  theory describing our four dimensional world
at low energies of the form
\be\label{4daction} S={1\over 2\k^2}\int\left (\sqrt{-G}R -\shalf F_4\w^* F_4\right )+S_m(G,\psi ),          \ee
where $S_m$ is the matter action and $F_4$ is a four form field strength satisfying the Bianchi identity $dF_4=0$. The equations of motion from this action are,
\brr\label{eofm}R_{\mu\nu}-\shalf G_{\mu\nu}R-{1\over 2.4!}(4F_{\mu...}F_{\nu}^{...}-\shalf G_{\mu\nu}F^2)&=&2\k^2T_{\mu\nu}\nn
d^*F&=&0.\err

In the above $T_{\mu\nu}$ is the matter stress tensor and we have ignored the 
matter equations of motion which will not play any role in this paper. Now the four form equation of motion and Bianchi identity have the solution,
\be\label{} F_4 =\mu^*1\ee
 where $\mu$ is a constant and the second factor  is the volume form. When this is substituted into the first equation one gets
\be\label{Rmunu} R_{\mu\nu}=G_{\mu\nu}(2\k^2V_0+\frac{\mu^2}{4})\ee

 Here $V_0$ is the effective cosmological constant generated in the matter sector.
Clearly if this is negative  then the integration constant $\mu$ can
always be chosen so as to get flat space. The question is what is the significance of this result. We should empahsize here that from the perspective of a four dimensional theory the integration constant $\mu$ can take any real value. However when the theory is embedded in a higher dimensional theory such as string theory which admits branes  which are sources of the four form field, $\mu$ is quantized and this mechanism would not work. We will encounter such a situation later on when we discuss what happens in string theoretic brane world scenarios.

First note that if one wants to argue that a flat space solution can
be obtained, even in the presence of quantum corrections to the matter action (ignoring gravity sector fluctuations), then one should replace
the classical matter action $S_m$ by the quantum effective action $\G_m$.  $V_0$ is now dependent on the RG scale and so  the integration constant needs to be renormalization scale dependent in 
order to get flat space at every scale. Of course such a  constant can be chosen at will but
to solve the cosmological constant problem \cite{weinberg} the question of  why out of the real line of values of this integration
constant, one particular value (or one value at each scale) 
gets chosen should also be answered. At present there is no clear answer to this and we will discuss this question further in section IV.

Nevertheless one may take the point of view that replacing a fine tuning problem with a choice of integration constant is progress, since one is not adjusting a parameter in the Lagrangian. In fact
in string theory there are no parameters to adjust and one might well
need a mechanism like this to get flat space solutions after supersymmetry breaking. So it might still be worth investigating
whether such mechanisms are available there. 

In the next section we will motivate a brane world scenario from a bottom up
approach as opposed to a top down string approach by asking  whether the RG scale in four dimensions can be thought of as a fifth dimension. In  section three we will discuss explicit embeddings of branes in five dimensions and discuss how the flat one and two brane solutions emerge. In fact in the string theory case we will argue
that the natural scenario is a two-brane one. From the five
dimensional point of view this requires a fine tuning of a parameter in the bulk potential, but we will argue that there are compactifications of string theory in which this parameter is (from the ten dimensional point of view) an integration constant. In the concluding section we discuss the problem of justifying the choice of integration constants
that leads to flat branes.
\section{Renormalization Group Flow in External Gravity.}
Let us consider the quantum theory corresponding to the classical action $S_m$. The fields $\psi$ could stand for the full set of standard model fields and we will also include a dilaton $\phi$ in order to make the connection later on to string theory. We are going to do semi-classical dilaton-gravity. In other words the dilaton gravity sector is treated classically while the standard model fields are treated quantum mechanically. The quantum theory is defined by the functional integral,
\be\label{quanttheory}e^{iW[G,\phi]}=\int [d\psi ]e^{iS_m[G,\phi, \psi ]}.\ee

In order to define  the quantum theory in a general gravitational background, a proper time cutoff propagator\cite{BD} 
\be K^{-1}_{\e} =\int_{\e_0}^{\e}e^{-Ks}ds\ee
is introduced with $K$ being the kinetic operator. Here $\e_0$ may be regarded as the ultra-violet cutoff (taken for instance to be the string scale) and $\e$ may either regarded as a renormalization scale or the scale defining a Wilsonian effective action. Using also the technique 
of Riemann normal coordinate expansions, one can derive in principle the quantum effective action in a systematic way preserving general covariance. 
The quantum action can therefore be written in a derivative expansion as
\be\label{quantaction} W[G, \phi ]= \int d^4x\sqrt{-G}(\Phi (\phi,\e )R -Z(\phi,\e )(\nabla\phi )^2 +V(\phi ) +...\ee
where the elipses represent higher derivative terms.   We have  indicated the explicit dependence on the RG scale. There would also of course be implicit dependence since the external fields $G, \phi$,like the couplings of the theory will aquire $\e$ dependence. Also we have set all expectation values of standard model fields to their values solving the equations of motion (at this point the functional $W$ is in fact equal to the 1PI effective action $\G$) and have been suppressed. 
The RG equation  reads,
\be\label{RG} \frac{dW}{d\e}=\frac{\pa W}{\pa\e}+\b_{\l}\frac{\pa W}{\pa\l}+\b_{\mu\nu}.\frac{\d W}{\d G_{\mu\nu}} +\b_{\phi}.\frac{\d W}{\d \phi}=0\ee
where the $\l$ are the couplings in the theory with associated beta
function $\b_{\l}$ and the other betas are the analogous beta functions for the metric and phi field (which are to be treated as generalized couplings). When the classical action for gravity and the $F_4$ field
are added to the above quantum action we again get an action of the form of (\ref{4daction}) (plus higher derivative terms) but with couplings which depend on $\phi$ and the RG scale $\e$. After a Weyl transformation this can be written as
\be\label{effaction} S={1\over 2\k^2(\e )}\int\left (\sqrt{-G}R -\tilde Z(\phi,\e )(\nabla\phi )^2-\shalf U(\phi,\e ) F_4\w^* F_4 -2\k^2V(\phi,\e )+...\right )       \ee
The previous argument still goes through with slight modifications.
For instance now the four form equation is replaced by
$d^*(U(\phi )F_4)=0$ which is solved by 
\be\label{Fsolution}F_4=\mu U^{-1*}1\ee 
(which also satisfies the Bianchi identity). But the main  result remains 
unchanged. The cosmological constant is an integration constant which
can be chosen (in a RG scale dependent way) so as to  get the effective cosmological constant to be zero.
The argument is robust under renormalization of the standard model
since it did not depend on particular functional forms of $Z, u$ or $V$. The problem of justifying the choice of integration constant
however remains.

Let us now ask the question under what circumstances can the  RG scale of the four dimensional theory be interpreted as a fifth dimension. In \cite{DVV} the argument was made that the five dimensional Hamilton-Jacobi equation can be interpreted as a four dimensional RG equation. Here we ask the opposite question; under
what conditions can the latter be interpreted as a five dimensional
gravity theory?

Consider the following expression constructed in terms of the quantum
effective action $W$ defined in (\ref{quantaction}),
\brr\label{HJ}{1\over \sqrt{- G}}{1\over 3}\left (G^{\mu\nu}\frac{\d W}{\d G^{\mu\nu}}\right )^2&-&\frac{\d W}{\d G^{\mu\nu}}\frac{\d W}{\d G_{\mu\nu}}-\shalf \left (\frac{\d W}{\d \phi}\right )^2\nn
&=&\sqrt{-G}(\tilde V(\phi,\e )+{1\over\k^2(\phi,\e )}R +M(\phi,\e )(\nabla\phi )^2 +...\err
The right hand side is just a consequence of general covariance
and the  ellipses stand for higher derivative terms. The particular form of the expression on the left hand side is of course chosen to agree with the corresponding expresion in the Hamilton-Jacobi equation of five dimensions \cite{DVV}. Under what conditions can $W$  be interpreted as a classical five dimensional action? Clearly this is possible
if the  explicit dependence on $\e$ is absent.\footnote{It should be noted that this explicit dependence includes the dependence on $\e$ through the renormalization of the flat space couplings as well. i.e. it corresponds to the first two terms of (\ref{RG}). } 
It is possible that this is the case in  ${\cal N}=4$ SU(N) Yang-Mills theory  (at least in the
large $N$ limit) and this  would then be an explanation of the AdS/CFT conjecture \cite{malda}.

Now the semi-classical theory of quantum fields is obtained after one adds a classical action and one then gets the action (\ref{effaction}). Let us set the $F_4$ terms to zero for the moment and ask what happens to the cosmological constant. Let $\phi =\phi_0(\e )$ be a constant field satisfying ${\pa V(\phi ,\e )\over\pa\phi} =0$. The gravity equation then  gives $R_{\mu\nu}=
\shalf V(\phi_0(\e ),\e )G_{\mu\nu}$. Clearly if the explicit dependence of
$V$ on $\e $ is absent then $\phi_0$ is $\e$ independent and so is the Ricci curvature so that if one has tuned the minimum of $V$ to zero at some scale (for instance $\e =\e_0$) then one will get flat space at all scales. But the issue is precisely for what theories in four dimensions is the statement of independence from $\e$  valid. With sufficient supersymmetry this could be the case. But  with ${\cal N}=1$ SUSY although the superpotential is not renormalized the Kahler potential is, so that the  potential for $\phi$ will in general depend explicitly on $\e$, though of course in this case one does not expect  renormalization of the minimum of the potential. Thus in order to have a
flat space solution at any RG scale one would in general need something like the mechanism discussed earlier.

Now it may be the case that, the absence of explicit dependence on 
$\e$ in $W$, while a sufficient condition for the five dimensional interpretation, is not be a necessary one. In other  words there could be a cancellation of the epsilon dependence on the LHS of
(\ref{HJ}) amongst the different terms so that the RHS is $\e$ independent. In this case just the 
mere fact that a five dimensional interpretation (as in the AdS/CFT case ) exists, is no gurantee of RG invariance of the four dimensional cosmological constant\footnote{Some discussion of the consequences of this are found in \cite{VV}.}. In other words the logic cannot be reversed. The absence of explicit dependence  of $W$  on  $\e$ (which implies in particular that the cosmological constant is not renormalized) is a sufficient condition for a  five dimensional interpretation, but the latter does not imply that the former is the case.

\section{Brane World Scenarios}

In the previous section we discussed the assumptions that would
lead us to interpret the RG scale as a fifth coordinate and thus
four dimensional semi-classical gravity as a five dimensional gravity theory. Here we will explicitly treat the four dimensional theory as
living on a brane in five dimensions. It is important to keep in mind the distinction between the two cases. In the first case the five dimensional theory (as for example in the AdS/CFT case) is simply a
dual representation of the four dimensional quantum effective action.
In the present case the underlying theory is five (or more dimensional) and the standard model is confined to a 3-brane living
in it. This may perhaps be realized in string theory as for example a type IIB  orientifold (compactified on some compact 5-manifold) with D3 branes and we shall discuss this further at the end of this section.

Using only general covariance, and keeping only  two derivative terms, the most general five dimensional action of gravity coupled to a scalar field is,
\be\label{fiveaction} S[G, \phi ]= \int d^5x\sqrt{-G}(R -Z(\phi,\e )(\nabla\phi )^2 +V(\phi ) +...)\ee
If this originates from the string theory example mentioned above, the 
potential $V$ may come from the $F_5$ terms that occur there, just like the $F_4$ terms in equation (\ref{effaction}), after using the solution to the equation of motion for the $F_5$ field\footnote{Thus in the notation of (\ref{effaction}) and the sentence below it, ( rewritten for five dimensions) $V$ in the above would be $\shalf\mu^2U(\phi )^{-1}$}.

Let us take the coordinates to be $x^M,~M=0,1,..4$ with the fifth coordinate $x^4=u$.
Now we insert 3-branes transverse to the direction $u$ at the points
$u=u_i$. We choose the static gauge so that the embedding functions
are $x^{\mu}(\xi )=\xi^{\mu},~\mu =0,..3$ and ignore their fluctuations. The effective action(s) coming from integrating the
``standard model" quantum fields (and hidden sector fields if there is more than one brane) will then take the form.
\be\label{brane} -\sum_i\int_{u=u_i}T_i(\phi ,\e )\sqrt{-G_{4(i)}}\ee
There will also be derivative terms but since we are interested in 
solutions with flat metrics and constant fields in 4d, they are irrelevant
to our discussion. The field equations for the system are then obtained by extremizing the sum of the two actions (\ref{fiveaction},\ref{brane}).

Now as in \cite{RS},\cite{Wise},\cite{deW} we look for solutions
that give flat space and constant $\phi$ field on the brane. So we write
\brr\label{fivemetric}ds^2&=& e^{2\o (u)}\eta_{\mu\nu}dx^{\mu}dx^{\nu}+du^2\nn
\phi &=& \phi (u).\err

The equations of motion then take the following form, (writing ${d\over du}\equiv '$)
 \brr\label{eofm} 6\o '^2&=&\shalf Z(\phi )\phi '^2 - V(\phi )\nn
3\o ''+6\o '^2 &=&-\shalf [Z(\phi )\phi '^2+V(\phi )]-\shalf\sum_i2\k^2T_i(\phi ,\e )\d (u-u_i)\nn
Z(\phi ''+4\o '\phi ' )+\shalf\phi '^2{d Z\over d\phi}&=&\shalf V'(\phi )+\k^2\sum_i{d T_i\over d\phi}\d (u-u_i)\err
The delta functions (due to the presence of the branes) imply that
$\o '$ and $\phi '$ are discontinuous at the branes and satisfy the matching conditions
\brr\label{match}3(\o '(u_i+0)-\o '(u_i-0 ))&=&-\k^2T_i|_{u_i}\nn
Z|_{u_i}(\phi '(u_i+0)-\phi '(u_i-0))&=&\k^2{dT_i\over d\phi}|_{u_i}\err
It should be noted that general covariance would imply that 
the scalar field equation should be satisfied when Einstein's equations (the first two in the above set (\ref{eofm})) are satisfied. In the presence of the branes (which break the five dimensional general covariance) the consistency of the third with the first two implies a 
condition
\be\label{consistency}(\phi '{dT_i\over d\phi})|_{u_i}=4(\o 'T_i)|_{u_i} \ee
where we may define $\phi '(u_i)=\shalf (\phi '(u_i+0)+\phi '(u_i-0))$ and similarly with $\o '(u_i)$. In fact this condition is the same as
what one would get from requiring that the potential be continuous at
$u=u_i$ and using the first equation of (\ref{eofm}). 
However when $\phi ' (0),\o ' (0)$ are zero (as is the case if we impose a $Z_2$ symmetry under $u\rightarrow -u$)  (\ref{consistency}) is trivially satisfied.

Let us first consider  solutions with one brane located say at
$u=0$. Also suppose that the bulk potential is of the form
\be\label{bulkpotential} V=\left ({\pa W\over \pa\phi }\right )^2-{4\over 3}W^2. \ee
where $W=W (\phi )$ may be considered to be a sort of superpotential. This form for $V$ arises naturally in gauged supergravities and appears to be a necessary condition for the existence of a solution
\cite{deW},\cite{DVV},\cite{townsend}. In this case the solutions  for the  warp factor and the scalar field can be obtained from \cite{deW},\cite{townsend},
\be  3\o '=-W(\phi ), ~~~~\phi '={dW\over d\phi}\ee
which can be solved by quadratures. Given these bulk solutions then
the existence of a flat  brane is guaranteed provided
the matching condition is satisfied. But this is just a matter of
choosing integration constants.

Let us discuss this further. 
We will impose a $Z_2$ symmetry as in \cite{RS},\cite{deW}. This might be a useful constraint in  that the most likely string realization of the brane world scenario is probably 
a type II orientifold. Thus we impose
\be\label{ztwo}\o (u)=\o (-u),~~~~\phi (u)=\phi (-u).\ee
The matching conditions (\ref{match}) become (for the brane at 
$u=0$),
\brr\label{match2} 3\o '|_{u=0+}&=&-\shalf\k^2T_0(\phi )|_{u=0+}\nn 
Z(\phi )\phi '|_{u=0+}&=&+\shalf\k^2{dT_0\over d\phi }(|_{u=0+}\err
The two second order equations for 
$\o$ and $\phi$ would have two integration constants each. However
the first equation of (\ref{eofm}) is an energy integral with the
total energy being zero. So the number of constants is reduced to three. Also a constant in $\o$ is irrelevant since the equations of 
motion do not involve $\o $ (this reflects the fact that such a constant can be absorbed in the rescaling of coordinates). Thus there 
really are only two constants (say $\phi (0)$ and $\o' (0)$) that can be then chosen to satisfy the
matching conditions. As explained in \cite{deW} when the first order equations in terms of
$W$ are being solved one would replace $\o '(0)$ by the integration constant coming from integrating (\ref{bulkpotential}). Thus with one
brane a flat solution can be obtained without any fine tuning. Such a one brane solution we believe is unlikely to arise say from string theory since the brane typically carries some charge which would mean that the fifth dimension would have to be non-compact. This may however be a way of getting the scenario of the second paper of \cite{RS}, but  with the exponential potential for $\phi$ that naturally arises in string theory, one gets a logarithmic  behaviour for the warp factor $\o $ \cite{Youm}\cite{ADS},\cite{KSS}\footnote{In the last two references the singularity in such a metric is interpreted as a point where the space is to be cut off. However it is not entirely clear to us how this can arise  from a microphysical theory such as string theory.} rather than the linear behaviour required in \cite{RS}. Later on we will come back to the
scenario of \cite{RS} in a situation where the modulus field has been integrated out from the low energy theory.

When there are two branes there is another pair of matching conditions to satisfy, but also there is another parameter namely the 
value  $u=R$ at which the new brane is situated (so in the IIB example this would be size of  the orbifolded  fifth dimension). Then we have an additional pair of conditions,
\brr 3\o '|_{u=R-}&=&+\shalf\k^2T_0(\phi )|_{u=R-}\nn 
Z(\phi )\phi '|_{u=R-}&=&-\shalf\k^2{dT_0\over d\phi }(|_{u=R-}\err
From the energy  constraint (the first equation of (\ref{eofm}) we also have
\be\label{tensionconst} Z^{-1}\left ({\k^2\over 2}{dT_0\over d\phi }\right )^2|_{u=0+}
-{4\over 3}\left ({\k^2\over 2}T_0\right )^2|_{u=0+}=V|_{u=0+}\ee
There is of course a similar equation at the point $u=R$ but this is not independent. Once we have a solution to the equations of motion
and the matching conditions this will be automatically satisfied.
In general the last equation will have a discrete set of solutions
for $\phi (0)$.

Thus there is one
extra condition and to satisfy that requires a fine tuning, either 
of the brane tension or of the potential \cite{deW}. 
However here
too fine tuning can be avoided if we make  at least one coupling constant in the potential  dynamical, i.e. an integration constant.
This is easily done if the bulk potential comes (at least partly) from the five dimensional analog of the $F^2$ term in (\ref{4daction}) or (\ref{effaction}). In this case, after solving the $F$ equation of motion as in the discussion in section I and substituting to get an effective action without $F$, one gets a 
potential for $\phi$ which depends on the integration constant $\mu$
as in the discussion after (\ref{4daction}). In the string theory context however (as we shall discuss in more detail later on) the brane is a source for the five form field and so $\mu$ is quantized. However in that case as we shall see later, if one considers (squashed) sphere compactifications one gets an additional adjustable parameter which is an integration constant of the ten dimensional theory. This is of course fine tuning from the  from the point of view of the five dimensional theory, but not from the fundamental ten dimensional point of view.

There are several different cases one may consider.\\
a) Supersymmetry is unbroken both in bulk and on brane(s).\\
b) Supersymmetry is preserved in bulk and broken on the brane(s).\\
c) Supersymmetry broken in both bulk and  brane(s).\\
d) Dilaton (and all other moduli) are fixed at the string scale

Let us discuss in turn the above cases.\\
a) Let us for example take a case which can come from type
IIB orientifold constructions compactified to five dimensions (say on a torus or an orbifold). The low energy effective action  contains  a term $\int U(\phi )F_5\w^* F_5$, where $\phi$ is the five dimensional dilaton. If one solves the equation of motion for the corresponding gauge field as in (\ref{Fsolution}) then one effectively gets a potential 
$V(\phi )=\mu^2 U^{-1}(\phi )$. Thus in the type IIB case,
$V =\mu^2exp
({5\over 3 }\phi )$.In this case $T_i =\tau_ie^{{5\over 6}\phi}$ where $\tau_i$ is the (constant) brane charge. If we substitute this in
 (\ref{tensionconst}) we see that in fact the $\phi (0)$ dependence drops out and the equation is satisfied if $\mu^2={13\over 9}({\k^2\over 2})^2\tau^2$. It should be noted that in this case even with one brane one needs the non-zero solution to the $F$ equation
(i.e. (\ref{Fsolution}). This is to be expected since as one crosses
the brane the $F$ field must change by the number of branes times the charge on a brane and in the supersymmetric case this charge is related to the tension.\footnote{This might be thought of as being
similar  to that discussed in \cite{BC} and \cite{KL}. In fact in the latter paper it is pointed out that this case does not exist in source free gauged supergravity. However our case is somewhat different in that we have explicit sources. So it is not completely clear to us that the arguments of \cite{KL} apply.} In the two brane case there is  no determination of the distance between the branes as is to be expected.

b) This case is more interesting. Now supersymmetry is broken on the brane and so the tension need not be as in a). In this case one would expect (\ref{tensionconst}) to determine $\phi (0)$ and the matching conditions will determine the other two integration constants. In the first order formalism  one of the 
integration constants will be the value of (say) $W(\phi (0))$. If we  work in the second order formalism  after fixing $\phi (0)$ as above the two constants to be determined by the the two matching conditions (\ref{match2}) are  $\phi '(0)$ and $\o '(0)$). Thus
one would indeed obtain (by choice of integration constants) a flat
brane in 4D without fine tuning. 
When there is a second brane however,
as we discussed earlier, there is one extra parameter (the distance $R$), but two more matching conditions to satisfy, and so we need to have
a dynamical bulk cosmological constant.

c) In this case the bulk potential will also get renormalized but as  far as the existence of flat brane solutions without fine tuning goes, there are no  qualitatively new features compared to b).

d) This case we believe is quite interesting since it seems very likely that the moduli are fixed at (or close to) the string scale.\footnote{For a discussion of this with references to earlier work see \cite{BdA}.} This as we mentioned earlier would  correspond to the original Randall-Sundrum scenario \cite{RS}.
This is possible in a  situation in which stringy non-perturbative effects
give a potential to all the moduli which should therefore be integrated out from the low energy effective action. In the absence of a string field theory, it is difficult to make precise statements 
. It is hard to see how the integration constant we want would arise
if all moduli are fixed at the string scale since all we know how to deal with are low energy actions. Nevertheless perhaps one can make some educated guesses about this case too.

A possible string theoretic construction for the scenarios in cases 
b) and c) and possibly d)  with  one modulus field (the breathing mode) remaining in the action, may run as follows. Consider type IIB  compactified to five dimensions on $S^5$ as in \cite{Bremer} (section 2.4). 
We assume that the ten dimensional dilaton $\vphi$ is fixed by string scale dynamics, but we keep the breathing mode that arises in the compactification, in the action.
The relevant part of the low energy effective action is 
\be\label{IIBeffective}S=\int d^5x\sqrt{-G}(R-(\nabla\phi)^2)-8m^2 e^{8\a\phi}+e^{{16\a\over 5}\phi}R_5)\ee
 In getting the above from the 10 D action  (or equations of motion) of IIB supergravity, the ansatz
$$ds_{10}^2=e^{2\a\phi}ds_5^2+e^{2\b\phi}ds^2(S^5)$$
with $\a ={1\over 4}\sqrt{5\over 3},~\b =-{3\over 5}\a$ has been made. $R_5$ is the Ricci scalar of the compact 5-manifold 
(in this case $S^5$) and is an unconstrained positive constant. The ansatz for the self-dual 5-form, is
\be\label{ffive}F_5 =4m e^{8\a\phi}\e_{(5)}+4m\e_{(5)}(S^5)\ee 
where $m$ is an integration constant.

At this point let us show out how the quantization of $m$ mentioned earlier comes about if this scenario emerges from string theory.
In this case the sources of the $F_5$  are $D_3$ branes (and/or orientifolds). The consistency of the the coupling of these objects
leads to the standard Dirac quantization rule,
$$\tau_3\int_{S_5}F_5=2\pi n$$
where $n$ is an integer. Using the ansatz (\ref{ffive}) for $F_5$
then gives $\tau_34m\pi^3r^5=2\pi n$ where $r$ is the radius of $S_5$.
This then gives us (after using the D3-brane tension formula $\tau_3=2\pi M_s^4$)
\be\label{m} m={n\over 4\pi^3M_s^4r^5}={nM_s\over 4\pi^3\hat r^5}\ee
where we have introduced
 the dimensionless parameter $\hat r\equiv rM_s$.
Also  the Ricci scalar of $S_5$ may then be written as $R_5={20M_s^2\over\hat r^2}$. Then the potential in (\ref{IIBeffective}) may be rewritten
\brr\label{potential} V&=&M_s^2\left [{n^2\over 2\pi^6\hat r^{10}}e^{8\a\phi}-{20\over \hat r^2}e^{{16\over 5}\a\phi}\right ]\nn
&=& M_s^2\hat r^{10\over 3}\left [{n^2\over 2\pi^6}e^{8\a\tilde\phi}-20e^{{16\over 5}\a\tilde\phi}\right ]\err 
where in the last line of the above equation we  put $\a\phi =\a\tilde\phi+{5\over 3}\ln\hat r$.

The minimum of this potential is given by $e^{{24\over 5}\a\tphi_0}={16\pi^6\over n^2}$.
As discussed in \cite{Bremer} the 5D action allows a $AdS_5$
solution with the cosmological constant being an integration constant, with  the breathing mode being fixed at ($\tphi_0$). The 
potential at this point gives a cosmological constant 
$\L = V(\tphi_0)=-(2\pi )^4{16\over n^2}M_s^2\hat r^{10\over 3}$. It should be noted that this five dimensional cosmological constant is
dependent  on the compactification parameter $\hat r$  and can be adjusted even though there are no adjustable constants in the 10 dimensional action.

Let us now compactify one of the spatial dimensions on $S^1/Z_2$ as before. In the original string theory scenario we would have
 $D3$ branes and orientifolds distributed over the five sphere but
we will just consider the effective theory in five dimensions with
two branes sitting at the ends of the compact fifth dimension giving us the scenario we had earlier.

The main focus of our discussion is going to be on how to get a two
-brane scenario without fine tuning of  parameters in the 10 dimensional Lagrangian. This will require that we keep the breathing mode $\tphi$ as a dynamical field that is not sitting at the minimum of the potential. However before we do that let us see whether we
can get a Randall-Sundrum \cite{RS} type scenario with the scalar breathing mode integrated out. For this we need to assume that some string non-perturbative effects fix the breathing mode at some high scale so that in the effective low energy five dimensional theory it has been 
integrated out just like the ten dimensional dilaton. Then we would have a bulk cosmological constant which will get contributions from
the string scale effects as well as from V. The latter will of course not necessarily be fixed at $V(\tphi_0)$ since it will be primarily determined by string effects however the important point is that such a contribution will be $\hat r$ dependent and hence will be adjustable. Making the mild assumption that this effective cosmological constant is negative we put   $V=-{\mu^2\over4}(\hat r)$ ) in the first equation of (\ref{eofm}). This then becomes  $ -12\o '^2=-{\mu^2\over 4}$
giving $\o ' =\pm {\mu\over 4\sqrt 3}$,
so that $\o =-{\mu\over 4\sqrt 3}|u|$.
In the last equation we have  used the $Z_2$ symmetry so as to obtain a warp factor that decays exponentially from the origin\cite{RS} on both sides. Using the matching condition (\ref{match2}) then gives
$\k^2 T_0={3\mu\over 2}$. The point is that this condition can be satisfied without fine tuning of the tension since the bulk potential is a function of the adjustable compactification constant $\hat r$.
If the size  of our compact direction is taken to infinity then we
have the RS2 scenario. On the other hand if this dimension is finite
then 
 we need a second brane at $u=R$  necessarily
its tension is negative $T_1=- {3\mu\over 2}$. This of course is
then a fine tuning.

There are several points that should be noted in this calculation. 
\begin{itemize}
\item In the absence of the modulus field there is no flat one brane
solution without fine tuning (as in \cite{RS}) or having a dynamical cosmological constant as in the above discussion. Indeed in the latter case there is (perhaps) a theory of confined gravity as in the second paper of \cite{RS} but obtained now without fine tuning of the fundamental (ten-dimensional) Lagrangian.
\item In the RS1 scenario the 
distance $R$ is now a free parameter (adjusted to a value that ``explains"  the gauge hierarchy in \cite{RS} ) and is not fixed by the dyanmics. Indeed the scalar field was introduced in \cite{Wise} in order to stabilize the value of $R$. However this requires a  tuning of a parameter in the potential in order to obtain the ``right" value. So unless this value of the parameter in the potential has a natural explanation there is no particular advantage to this. 
\item In the two brane case the so-called visible brane (on which the standard model is supposed to live) has negative tension. Also since the dynamical bulk
cosmological constant tracks the brane tension at the origin as it changes with RG scale the only way (without fine tuning) for a two
brane solution to be viable is for the RG flow of the visible brane
to be the same in magnitude though opposite in sign as on the other brane. It is not clear to us how to achieve this in a natural way.
Thus this still requires fine tuning.
\end{itemize}

Let us now get back to the main point of this discussion. To discuss a two brane case (in cases b),d) above) with at least one modulus field dynamical we should look at a IIB orientifold.
Now the fifth dimension is an interval $S_1/Z_2$ with 16 orientifold fixed planes at the fixed points $u=0$ and $u=R$. One also needs to introduce D-branes in order to cancel tadpoles\footnote{Indeed such a model is T-dual to the type IA theory discussed in \cite{pw}.}. 
In the presence of D-branes and orientifold planes that are charged under the $F_5$ field we have (as in \cite{pw}) a discontinuity in 
$m$ by an amount equal to the brane charge/tension at the position of the brane. Imposing also the $Z_2$ symmetry  the constant $m$ in (\ref{IIBeffective}) would be fixed in terms of the brane charge as in \cite{pw}. This is of course just another way of verifying the quantization discussed earlier. When supersymmetry is broken however the brane tension would get renormalized so that the supersymmetric relation between tension and charge will be lost. Nevertheless the  constant $\hat r$   can adjust itself now to track the brane tension.
In addition (assuming it is not fixed at the string scale by stringy effects as in the previous example) we have a modulus field $\phi$ as in the  discussion at the begining of this section, to supply an addtional integration constant so that 
one may have solutions with two flat branes.

Let us give some more details of this scenario. The effective five dimensional theory in the presence of branes is given by (after putting in also the value of the five D Newton constant)
\brr\label{action2} S&=&2\pi^4M_s^3\hat r^5\int d^5x\sqrt{g_5}
\left (R-\shalf(\pa\tphi)^2+M_s^2\hat r^{10\over3}U(\tphi )\right )\nn &+&{2\pi M_s^4\over\pi^3}\left\{\int_{u=0}d^4x\sqrt{g_4}T_{0}(\tphi )\int_{u=R}d^4x\sqrt{g_4}T_{1}(\tphi )\right\}.\err
In the above we have written the potential $V = M_s^2R^{10\over 3}U(\tphi)$ where $U$ is independent of $\hat r$ (see (\ref{potential})). Let us first look at what might happen in the supersymmetric case. Here in analogy with the case discussed by Lucas
et al \cite{lucas} for the Supersymmetry preserving compactification of M theory on ${S^1\over Z_2}\times CY_3$ we expect the BPS equations $$T_0(\tphi )=-T_1(\tphi )=\pi^3\hat r^{5/3}W(\tphi )$$ where
$$U(\tphi )=\left ({\pa W\over \pa\tphi }\right )^2-{2\over 3}W^2$$.
(Note that  apart from normalization this is the same $W$ as before).
In fact these BPS conditions have been justified for this system very recently in \cite{bkv},\cite{dls}. In this case it is easily seen that the parameter $\hat r$ drops out of the matching conditions which however can be satisfied for arbitrary inter-brane distance $R$
and integration constants. This is exactly like what happens in the
M theory case investigated by Lukas et al. \cite{lukas}.

In the broken supersymmetric case however the situation is quite different. For instance we may imagine now is that the compactification is on a squashed sphere so that the supersymmetry is  $N=1$ which is then broken by for instance by gaugino condensation effects on one or other brane. Now the tension will be renormalized
from its BPS value so that one would expect
$$T_i(\tphi )\rightarrow\pm\pi^3\hat r^{5\over 3}W(\tphi )+\e\psi_i(\hat r, \tphi )$$

Here the function $\psi$ is dependent on the details of the low energy field theories on the branes. Thus now $\hat r$ will not drop out of the matching conditions and indeed we need to adjust the two integration constants $\tphi (0), A'(0)$ the inter-brane distance $R$
as well as the compactification parameter $\hat r$ in order to get 
a flat four-dimensional brane world.  

A detailed discussion of such models will be given in a forthcoming paper \cite{AFI}.
\section{Conclusions}
Let us first discuss the results of \cite{ADS},\cite{KSS}. From our discussion it should be clear that the reason that flat (one) brane solutions are obtained (without fine tuning) in these works is that 
integration constants have been chosen to ensure the existence of such solutions. Of course since these authors do not discuss two brane 
solutions  they do not need the $F_5$ field or the Ricci non-flat compactification that we have introduced. However as we argued (and is indeed implied by the work of de Wolf et al \cite{deW}) one flat brane solution is obtained 
in the presence of a dynamical scalar field by choosing the integration constant $\phi (0)$ appropriately. It does not depend on the particular form of the brane tension $T(\phi )$ as seems to be implied in \cite{ADS}. Indeed this is just as well since the form of this function can change under renormalization effects on the brane.
The fact that {\it only} a flat brane is allowed for a particular
form of this function (see equation (14) of \cite{ADS} and section (3.2)
of \cite{KSS2}) therefore is  not a RG invariant statement. Quantum effects of the standard model in a background metric yields both a cosmological constant  as well as curvature terms (as in our (\ref{quantaction})). The latter will necessarily modify these arguments. 

The main conclusion of the present work is that one can indeed
obtain flat branes (and in particular zero cosmological constant in the   brane containing the standard model) without fine tuning, but
it involves a choice of integration constants/compactification parmeters. In this respect these
theories have the same problem that bedevilled those of references
\cite{DvN},\cite{Hawking} \cite{Witten},\cite{Rub2}. It is useful to review this issue briefly. The point is to show that the particular integration constant(s) that leads to a zero cosmological constant gets chosen because it is the most probable one. To show this Hawking used a Euclidean quantum gravity argument according to which (see also section VIII of Weinberg's classic review \cite{weinberg})  the probability for the occurence of a value $\mu$ for the integration constant was given by $P(\mu)\propto \exp (-\G_E [\psi_c])$ where $\G_E$ is the Euclidean quantum effective action (essentially our equation (\ref{effaction}) Wick rotated to  a Euclidean metric) and the $\psi_c$ are the values of all the fields 
evaluated at an extremum of $\G$. The Euclidean (effective) action for a $D$ dimensional theory after setting all  other fields but the metric to their 
quantum ground state values as above would take the form, (setting the $D$ dimensional Planck mass equal to one)
$ \G_E = -\int \sqrt G (R-\L  )$. From the Einstein equation we have
$R={D\L\over (D-2)}$. Substituting this into the Euclidean action 
gives $S_E=-{2V_D\over D-2}\L$ where $V_D$ is the volume of Euclidean
D space. If $\L$ is positive then the space is $S_D$ and its volume
is $V_D={a^2\over \L^2}$ so that the action becomes $S_E=-{2V\over (D-2)\L}$. Thus the probability distribution becomes 
\be\label{prob}P(\mu )\propto  e^{-\G_E [\psi_c ]}=e^ {+{2V\over (D-2)\L}}.\ee 
This would 
 imply that the probability was peaked at $\L \rightarrow 0+$.
\footnote{It was pointed out by Duff \cite{Duff} that if one substitutes the solution for
$F$  into Einstein's equation and then infers the effective action from which it comes, then one finds in fact that the (Euclidean) action is positive near $\L = 0$ so that this value is actually disfavoured! Nevertheless it was claimed in \cite{DJ} that the correct action
has a boundary term that does not affect the equation of motion and
its inclusion will reestablish Hawking's result. I wish to thank R. Bousso for pointing out this reference to me. On the other hand as pointed out by M. Duff (private communication) Hawking's Euclidean space no-boundary action should not of course have a boundary contribution! The situation therefore remains murky.}

In our case it is not clear whether an analog of Hawking's argument would work\footnote{Hawking's argument may work also in the case of unimodular gravity \cite{Ng2}.}. However one would think that one should apply the argument to the five (or ten?) dimensional theory since that is the action one is starting from. But the integration constants must get chosen so that it is the four dimensional theory that has to have zero cosmological constant. At this point it is not clear to us whether a version of this argument can be used to justify the choice of integration constants.    

Note added: While this paper was being prepared for submission,
a paper which (inter alia) makes comments related to ours about the one brane case
of \cite{ADS},\cite{KSS}, appeared as an e-print \cite{gubser}.

\section{Acknowledgements}I wish to thank Alex Flournoy and Nicos Irges for discussions and Shamit Kachru for correspondence on the work of \cite{ADS},\cite{KSS}. I'm also grateful to Jack Ng and Paul Townsend for setting me straight on the historical record, and Renata Kallosh for correspondence on \cite{KL}. This work is partially supported by the Department of
Energy contract No. DE-FG02-91-ER-40672.


\end{document}